\documentclass[nohyperref]{article}

\usepackage{microtype}
\usepackage{graphicx}
\usepackage{subcaption}
\usepackage{multicol}
\usepackage{booktabs} 

\usepackage{hyperref}

\usepackage[accepted]{ai4science}

\usepackage{amsmath}
\usepackage{amssymb}
\usepackage{mathtools}
\usepackage{amsthm}

\usepackage[capitalize,noabbrev]{cleveref}

\theoremstyle{plain}

\theoremstyle{definition}

\theoremstyle{remark}

\usepackage[textsize=tiny]{todonotes}

\icmltitlerunning{Unsupervised Discovery of Inertial-Fusion-Physics using AD + a PDE Solver + a Maximum Entropy Loss Function}

\begin{document}

\twocolumn[
\icmltitle{Unsupervised Discovery of Inertial-Fusion Plasma Physics using Differentiable Kinetic Simulations and a Maximum Entropy Loss Function}



\icmlsetsymbol{equal}{*}

\begin{icmlauthorlist}
\icmlauthor{Archis Joglekar}{ergo,umich} 
\icmlauthor{Alexander Thomas}{umich}
\end{icmlauthorlist}

\icmlaffiliation{ergo}{Ergodic LLC, San Francisco, CA, USA}
\icmlaffiliation{umich}{University of Michigan, Ann Arbor, MI, USA}

\icmlcorrespondingauthor{Archis Joglekar}{archisj@umich.edu}

\icmlkeywords{Machine Learning, ICML}

\vskip 0.3in
]



\printAffiliationsAndNotice{}  

\begin{abstract}
Plasma supports collective modes and particle-wave interactions that leads to complex behavior in inertial fusion energy applications. While plasma can sometimes be modeled as a charged fluid, a kinetic description is useful towards the study of nonlinear effects in the higher dimensional momentum-position phase-space that describes the full complexity of plasma dynamics. We create a differentiable solver for the plasma kinetics 3D partial-differential-equation and introduce a domain-specific objective function. Using this framework, we perform gradient-based optimization of neural networks that provide forcing function parameters to the differentiable solver given a set of initial conditions. We apply this to an inertial-fusion relevant configuration and find that the optimization process exploits a novel physical effect that has previously remained undiscovered.
\end{abstract}

\section{Introduction}
\label{intro}
\subsection{Fusion, Charged Fluid Dynamics, and Kinetics}
Thermonuclear fusion involves conditions where matter is best represented as a warm plasma where the ions and electrons form a charged particle soup. Such plasmas can often be described using a combination of Maxwell's equations and the Navier-Stokes equations. However, many non-linear effects in plasmas cannot be modeled using the fluid description without further modifications.

These effects are termed kinetic effects. To model kinetic effects, a more complete description  is required. This can be accomplished by either retaining the velocities of individual particles of the fluid or retaining a statistical representation of the particles. In either case, the objective becomes to model the phase space dynamics rather than a version integrated in velocity.

Phase space dynamics for plasmas are given by formulations or realizations of the non-linear transport equation. A common simplification that is valid for fast time-scales, is to assume the ions, and their phase space, remains stationary. Then, the transport equation for electrons is given by
\begin{equation}
    \frac{\partial f}{\partial t} + v\frac{\partial f}{\partial x} - E \frac{\partial f}{\partial v} = \left[\frac{\delta f}{\delta t}\right]_{coll} \label{eq:vfp},
\end{equation}
where $E = \int f ~dv - 1$, and $f = f(t,x,v)$. Equation \ref{eq:vfp}, along with Gauss's Law, is often termed the Vlasov-Poisson-Fokker-Planck (VPFP) equation set. Solving the VPFP set is often analytically intractable, even in one spatial and one velocity dimension (1D-1V). This is because the left-hand-side has a stiff linear transport term, has a non-linear term in $E \partial f/\partial v$, and can sustain wave propagation and other hyperbolic partial-differential-equation (PDE) behavior. Additionally, the right hand side is typically represented by a hyperbolic, advection-diffusion, partial-differential-equation. Making progress on kinetic plasma physics requires computational simulation tools.

 Numerical solutions to the 1D-1V VPFP equation set have been applied in research on laser-plasma interactions in the context of inertial fusion  plasma-based accelerators \cite{thomas_vlasov_2016}, space physics \cite{chen_evidence_2019}, fundamental plasma physics \cite{pezzi_vida_2019}, and inertial fusion \cite{Strozzi2007, fahlen_propagation_2009, banks_two-dimensional_2011}

In this work, we extend previous findings \cite{fahlen_propagation_2009} in the context of inertial fusion using a gradient-based approach towards discovery of new physics. 

\begin{figure*}[ht]
\centering
    \includegraphics[width=\textwidth]{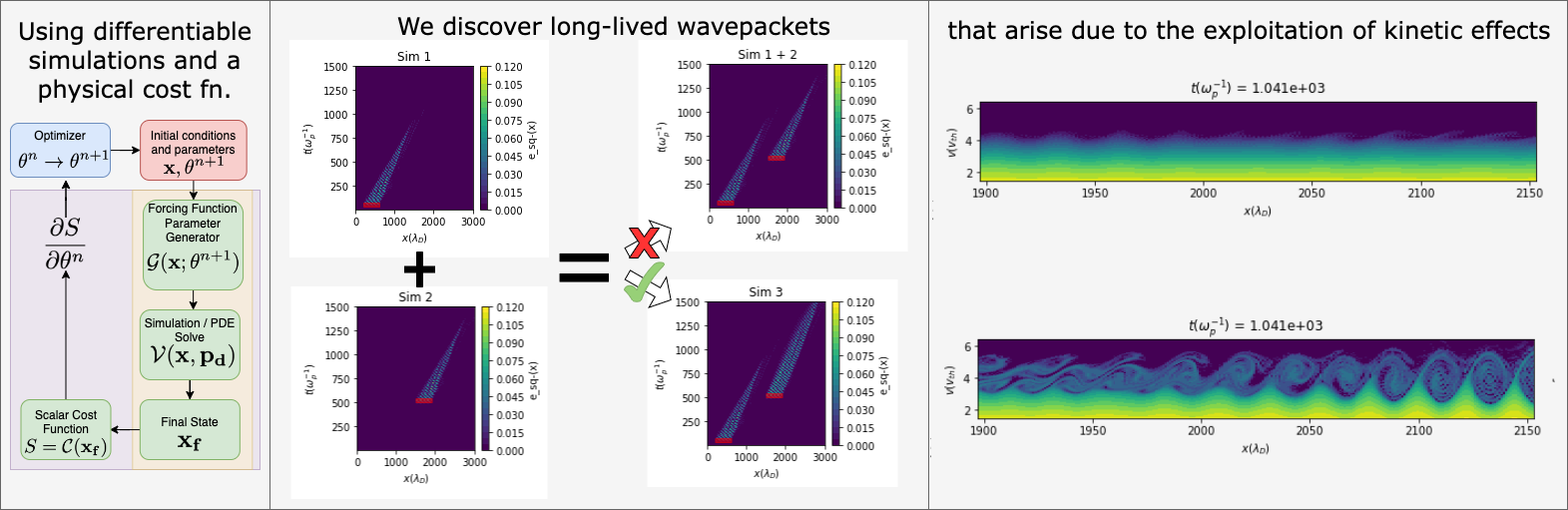}
    \caption{\textbf{Left} - We create a differentiable loop to learn forcing function parameters that optimize for a free energy and entropy metric. \textbf{Middle} - The optimization uncovers a superadditive process that results in a long-lived second wavepacket. \textbf{Right} - Deeper inspection shows that the learning process exploits novel kinetic plasma physics in order to minimize damping, where the second wavepacket interacts with the electrons from the first wavepacket.}
    \label{fig:mainfig}
\end{figure*}

\subsection{Gradient-Based Optimization in Plasma Physics}
There have been several recent applications of gradient-descent towards fusion plasma physics. Analytic approaches have resulted in the development of adjoint methods for shape derivatives of functions that depend on magnetohydrodynamics (MHD) equilibria \cite{antonsen_adjoint_2019, paul_adjoint_2020}. These methods have been used to perform optimization of stellarator design \cite{paul_gradient-based_2021}. In other work, by using gradients obtained from analytic \cite{zhu_new_2017} and automatic differentiation (AD) \cite{mcgreivy_optimized_2021}, the FOCUS and FOCUSADD codes optimize coil shape. These advances are founded on the concept of performing a sensitivity analysis towards device design. 

This work applies AD towards learning physical relationships and discovering novel phenomena in the VPFP dynamical system. We do this by training neural networks through differentiable simulations that solve the 1D-1V version of eq. \ref{eq:vfp} and optimize for a custom objective function. 

\subsection{Applications of Differentiable Physical Simulations}
Differentiable physics simulations have been used in a variety of contexts. For example, for learning parameters for molecular dynamics \cite{schoenholz_jax_2019}, for learning differencing stencils in PDEs \cite{bar-sinai_learning_2019, zhuang_learned_2020, kochkov_machine_2021}, and for controlling PDEs \cite{holl_learning_2020}.

In the work here, we train a neural network that provides control parameters to the PDE solver. We choose physical parameters as inputs and control parameters as outputs of the neural network. This enables the neural network to learn a function that describes the physical relationship between the plasma parameters and the forcing function parameters e.g. the resonance frequency. 

We train the neural network in an unsupervised fashion using a cost function based on the maximum entropy principle. This enables us to create self-learning plasma physics simulations, where the optimization process provides a physically interpretable function that can enable physics discovery.

\section{Parameter and Function Learning using Differentiable Simulations}
\begin{figure*}[ht]
\begin{subfigure}{0.22\textwidth}
\includegraphics[width=\textwidth]{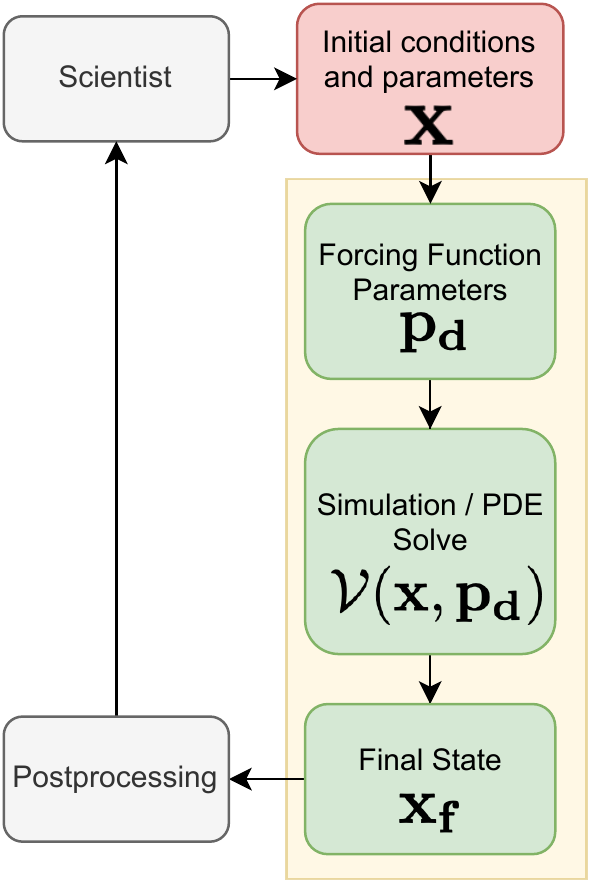} 
\caption{Open Loop - Manual Workflow}
\label{fig:basic_sim}
\end{subfigure}
\hfill
\begin{subfigure}{0.22\textwidth}
\includegraphics[width=\textwidth]{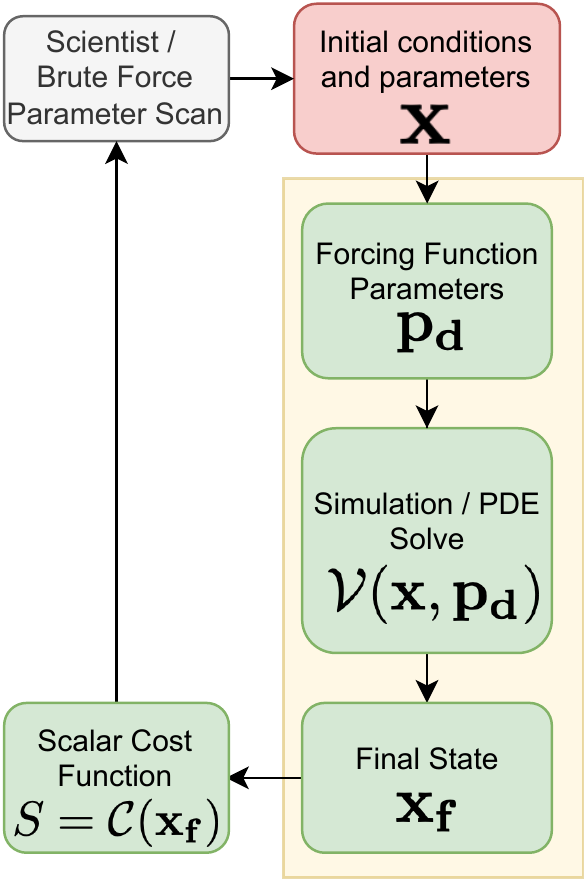} 
\caption{Closed Loop - Brute Force Parameter Scan}
\label{fig:param_scan}
\end{subfigure}
\hfill
\begin{subfigure}{0.22\textwidth}
\includegraphics[width=\textwidth]{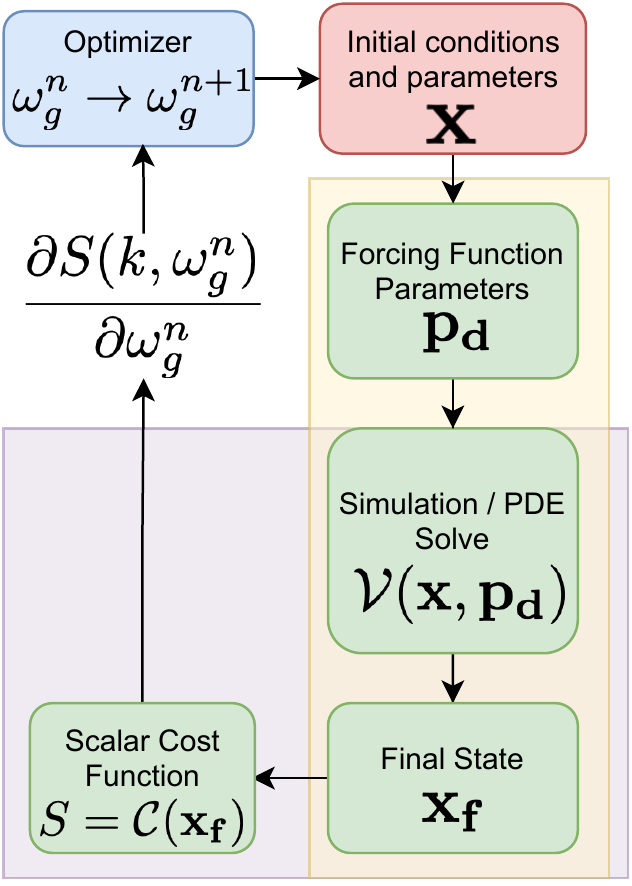} 
\caption{Gradient-Descent-Based Parameter Learning}
\label{fig:param}
\end{subfigure}
\hfill
\begin{subfigure}{0.22\textwidth}
\includegraphics[width=\textwidth]{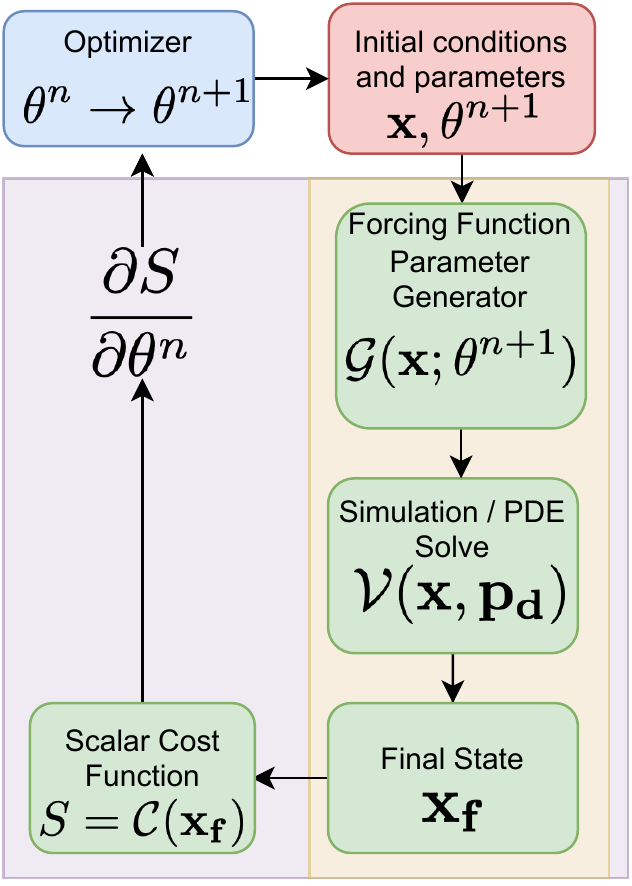} 
\caption{Gradient-Descent-Based Function Learning}
\label{fig:nn}
\end{subfigure}

\caption{(a) A typical workflow of a computational scientist where the scientist provides the initial conditions and forcing function parameters to a partial-differential-equation (PDE) solve. The output of the solve is stored as the final state $\mathbf{x_f}$. The final state is analyzed by the scientist using domain-specific postprocessing algorithms. (b) A cost function and a parameter scan is introduced which enables a closed-loop search. (c) A gradient-descent based optimization algorithm replaces the parameter-scan to provide a more efficient search mechanism. This requires the components in the purple background to be written in an auto-differentiable framework (d) We add a neural network that generates the forcing function parameters as a function of other parameters. This generalizes the learnings from (c) and enables interpolation and extrapolation within the learned parameters.}
\label{fig:workflows}
\end{figure*}
In this section, we provide a step-by-step description of how a traditional simulation-based computational physics workflow may be modified to perform closed-loop optimization as applied to parameter and function learning. 

\subsection{Open Loop - Manual Workflow}
Figure \ref{fig:basic_sim} depicts a typical workflow of a computational scientist represented as a cyclic graph. The scientist defines the parametric inputs which create the state vector $\mathbf{x}$. This can contain any parameters that are used to define the simulation e.g. the grid size, the number of solver steps, etc. For didactic purposes, the physical parameters to the simulation may be separated into a different vector of inputs $\mathbf{p_d}$ e.g. the forcing function parameters, the viscosity coefficient etc.

Each of $\mathbf{x}$ and $\mathbf{p_d}$ is passed to the algorithm that solves the PDE which is represented by the function, $\mathcal{V}$. The output of these simulations is stored in the final state vector $\mathbf{x_f}$. The final state is postprocessed using a domain-specific set of algorithms devised by the scientist or otherwise. The results of the postprocessing are interpreted by the scientist who then determines the next set of inputs and parameters.


\subsection{Closed Loop - Brute Force Parameter Scan}
Figure \ref{fig:param_scan} shows a more automated workflow. We replace the gray-box postprocessing step with the calculation of a scalar quantity $S$ using a Cost Function $\mathcal{C}$ on the final state $\mathbf{x_f}$. This reduces the complexity of the interpretation of the postprocessing and enables a more rapid search in parameter space. The decrease in required human effort for completing one cycle enables the scientist to execute this loop as a brute force parameter scan over a pre-defined parameter space. At the end, the scientist can look up the minimum/maximum of the scalar cost function, and find the parameters which provide that minimum. 

The parameter scan approach scales with the number of different unique parameters and the number of values of each parameter. e.g. a 2-D search in $x$ and $y$ requires $N_x \times N_y$ calculations. Therefore, the parameter scan approach quickly becomes inefficient when there are many parameters to scan, or when the required resolution in parameter space is very high. To search this parameter space efficiently, and to escape the linear scaling with each parameter, we can use gradient descent. 

\subsection{Gradient-Descent-Based Parameter Learning}
\label{sec:gbpl}
Figure \ref{fig:param} includes two modifications. The scientist/parameter search graybox has been replaced with a gradient-descent-based optimization algorithm. This algorithm provides the updated parameters, e.g. $\omega_G$, a guess for the resonant frequency of the system, for the next iteration of the loop. The gradient-descent algorithm requires the calculation of an accurate gradient. By writing our PDE solver $\mathcal{V}$ and the cost function $\mathcal{C}$ using a numerical framework that supports automatic differentiation, we are able to calculate $\partial S / \partial \omega_G$. 

Since
$$
    \mathcal{S} = \mathcal{C}(\mathbf{x_f}) = \mathcal{C}(\mathcal{V}(\mathbf{x}, \mathbf{p_d})),
$$ the gradient for the update-step is given by
$$
    \frac{\partial \mathcal{S}}{\partial \mathbf{p_d}} = \frac{\partial \mathcal{C}(\mathcal{V}(\mathbf{x}, \mathbf{p_d}))}{\partial \mathbf{p_d}} = \frac{\partial \mathcal{S}}{\partial \mathcal{V}} \frac{\partial \mathcal{V}}{\partial \mathbf{p_d}}.
$$

For example, if we wish to learn the resonant frequency, $\omega$, that optimizes for the scalar, $\mathcal{S}$, we compute 
\begin{align}
    \frac{\partial \mathcal{S}}{\partial \mathbf{\omega}} =  \frac{\partial \mathcal{S}}{\partial \mathcal{V}} \frac{\partial \mathcal{V}}{\partial \mathbf{\omega}}. \label{eq:gbpl-grad}
\end{align}

Assuming a well-behaved solution manifold, performing gradient-descent tends to reduce the number of iterations required to find the minimum in comparison to a evenly-spaced parameter scan, especially when the required resolution is unknown \cite{NoceWrig06}. Put another way, gradient-descent effectively provides an adaptive stepping mechanism in-lieu of the pre-defined values that represent a parameter scan.

\subsection{Gradient-Descent-Based Function Learning}
In the final step, shown in fig. \ref{fig:nn}, we can replace the lookup-like capability of the parameter optimization and choose to learn a function that can provide the learned parameters. The advantage to learning a function, however, is the ability to interpolate and extrapolate with respect in the input space. 

Here, we choose to use neural networks, with a parameter vector $\theta$, to parameterize the desired function. This allows us to extend the gradient-descent based methodology and leverage existing numerical software to implement this differentiable programming loop.

Now,
$$
    \mathcal{S} = \mathcal{C}(\mathbf{x_f}) = \mathcal{C}(\mathcal{V}(\mathbf{x}, \mathbf{p_d})) = \mathcal{C}(\mathcal{V}(\mathbf{x}, \mathcal{G}(\mathbf{x}, \mathbf{p_d}; \theta)),
$$ where $\mathcal{G}$ is a function that generates the desired forcing function parameter given a parameter vector $\mathbf{\theta}$. To extend the example from sec. \ref{sec:gbpl}, $\omega$ is now a function given by $\omega = \mathcal{G}(\mathbf{x}, \mathbf{p_d}; \theta)$. 

We compute the same gradient as in eq. \ref{eq:gbpl-grad} and add a correction factor that arises because the parameter (vector) is now $\theta$, rather than $\omega$. The necessary gradient for the gradient update is now given by
\begin{align}
    \frac{\partial \mathcal{S}}{\partial \mathbf{\theta}} = \left[\frac{\partial \mathcal{S}}{\partial \mathcal{V}} \frac{\partial \mathcal{V}}{\partial \mathcal{G}} \right] \frac{\partial \mathcal{G}(\mathbf{x}, \mathbf{p_d}; \mathbf{\theta})}{\partial \mathbf{\theta}}. \label{eq:nnloss}
\end{align}

\section{Physics Discovery as an Optimization Problem}
We will start with a basic physical effect that is well understood, Electrostatic Wavepackets in Inertial Fusion Plasmas. Then we will ask the question of "what happens when you excite this system?"

\subsection{Electrostatic Wavepackets in Inertial Fusion}
When electrostatic waves are driven to large amplitude, electrons can become trapped in the large potential \cite{oneil_collisionless_1965}. Simulations of Stimulated Raman Scattering (SRS) in inertial confinement fusion (ICF) scenarios show that large-amplitude waves of finite extent are generated in the laser-plasma interaction, and that particle trapping is correlated with the transition to the high-reflectivity burst regime of SRS  \cite{Strozzi2007, ellis_convective_2012}. 

Simulating wavepackets, similar to those generated in SRS, but in isolation, has illuminated kinetic dynamics where trapped electrons, with velocity $\approx v_{ph}$, transit the wavepacket which is moving at the slower group velocity $v_g$. The transit of the trapped electrons from the back of the wavepacket to the front results in the resumption of Landau damping at the back and the wavepacket is then damped away \cite{fahlen_propagation_2009}. 

Recent work modeled the interaction of multiple speckles with a magnetic field acting as a control parameter. Since the effect of the magnetic field is to rotate the distribution in velocity space, the field strength serves as a parameter by which the authors control scattered particle propagation. Using this, along with carefully placed laser speckles, they show that scattered light and particles can serve as the trigger for SRS \cite{winjum_interactions_2019}.

Here, we ask \emph{What happens when a non-linear electron plasma wavepacket is driven on top of another?} 

To answer this question, we reframe it as an optimization problem and ask, \textbf{What is the best way to excite a wavepacket that interacts with a pre-existing wavepacket?}

To solve this optimization problem, we turn to the framework established in the previous section and in fig. \ref{fig:workflows}. To automate this process, we can use the workflow described in fig. \ref{fig:param} or fig. \ref{fig:nn}. In both, the physicist is still required to provide their expertise in the form of 1/ the parameterization, and 2/ the objective function.

\subsection{Physicist Input 1: Parameterizing the Forcing Function with a Neural Network}
\begin{figure}
    \centering
    \includegraphics[width=0.4\textwidth]{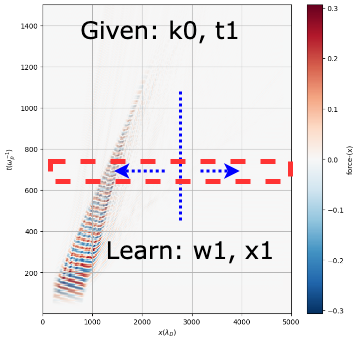}
    \caption{Given a first wavepacket with wavenumber $k_0$ and a desired time of second wavepacket excitation $t_1$, the task is to learn functions that give the optimal frequency $\omega_1$ and spatial location $x_1$ of the second wavepacket.}
    \label{fig:wavepacketdiscovery}
\end{figure}
We start with a large-amplitude, finite-length electrostatic wavepacket driven by a forcing function with parameters given by 
\begin{align}
    \mathbf{p_0} &= \left[x_{0}, \omega_{0},t_{0},k_{0}\right],
\end{align}
where $x_{i}$ is the location of excitation, $\omega_i$ is the frequency, $t_i$ is the time of excitation, and $k_i$ is the wavenumber, of the $i^\text{th}$ wavepacket.

Since we seek to excite a second wavepacket that can interact with the detrapped electrons, we set the wavenumber $k_1 = k_0$. We reparameterize the resonant frequency, $\omega_1$, with a frequency shift, $\Delta \omega_1$ and the linear resonant frequency $\omega_0$ such that $\omega_1 = \omega_0 + \Delta \omega_1$. 

We use the time of excitation of the second wavepacket, $t_1$, as an independent variable along with $k_0$. For each $t_1$ and $k_0$, we seek to learn functions that produce $x_1$ and $\Delta \omega_1$ i.e. we seek to learn $x_1(t_1, k_0)$ and $\Delta \omega_1( t_1, k_0)$. The entire parameter vector for the second wavepacket is given by 
\begin{align}
    \mathbf{p_0} &= \left[ x_{1}(t_1, k_0), \Delta \omega_{1}(t_1, k_0),t_{1}, k_{0} \right].
\end{align}
This framing is also illustrated in fig. \ref{fig:wavepacketdiscovery} where given $k_0$ and $t_1$, we seek functions for $\omega_1$ and $x_1$.

\subsection{Physicist Input 2: A Minimum Energy and Maximum Entropy Loss Function}
We reparameterize $\Delta \omega_1$ and $x_1$ with a neural network with a parameter vector, $\theta^*$, that maximizes the electrostatic energy (minimizes the free energy) and maximizes the kinetic entropy i.e.

\begin{align}
    x_1 &= x_1(t_1,k_0; \theta^*), \\
    \Delta \omega_1 &= \Delta \omega_1(t_1, k_0; \theta^*).
\end{align}
where,
\begin{align}
    \theta^* &= argmin ~~\left[-E_\text{es}(\textbf{p}; \theta) -\Delta \mathcal{KE}(\textbf{p}; \theta)\right], 
\end{align}
and
\begin{align}
    E_\text{es} &= \sum_{t_{i}}^{t_f} \Delta t \sum_x \Delta x ~ E_x^2 \\
    \Delta \mathcal{KE} &= \sum_{t_{i}}^{t_f} \Delta t \sum_x \Delta x \sum_v \Delta v ~ (f \log(f) - f_\text{MX} \log(f_\text{MX})),
\end{align}
are the electrostatic energy, and entropy, terms in the loss function, respectively. $f_\text{MX} = f_\text{MX}(n, T), n = \int f dv, T = \int f v^2 dv$ where $f_\text{MX}$ is the Maxwell-Boltzmann distribution.

\section{Function Learning Results in the Exploitation of Novel Physics}
In this section, we describe the outcome of posing "Physics Discovery as an Optimization Problem" using our differentiable simulations framework. We start at the highest level, describing the training dynamics, and work our way deeper into the results by attempting to interpret what the training process has learned.

\subsection{Training Dynamics suggest Sample Efficiency}
\begin{figure}[h]
    \centering
    \includegraphics[width=0.45\textwidth]{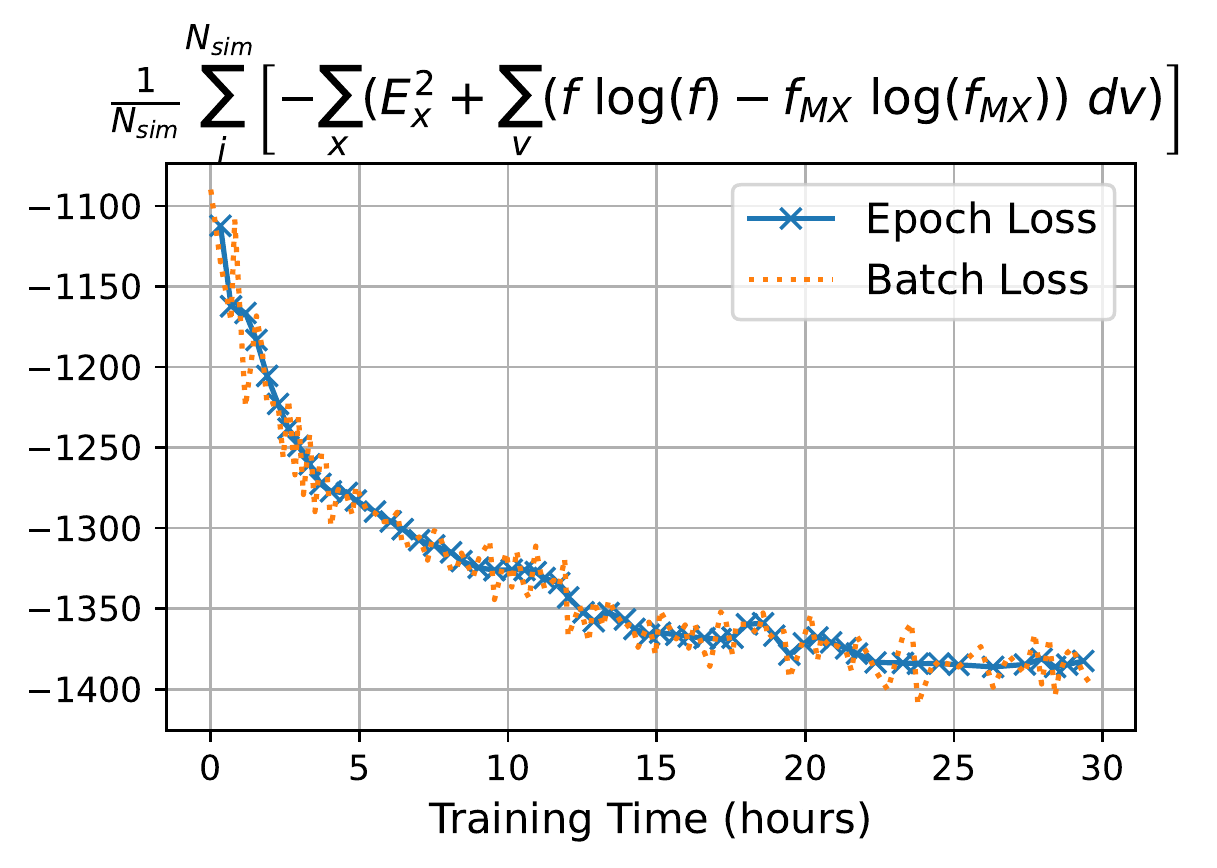}
    \caption{The loss given by eq. \ref{eq:loss} is plotted as a function of time. Each cross represents an epoch, and batch-wise fluctuations are also displayed. The training converges after roughly 30 hours}
    \label{fig:loss}
\end{figure}

Figure \ref{fig:loss} shows that the loss value is reduced over the duration of the training process, roughly 30 hours, which runs for 60 epochs. The convergence in the loss metric suggests that we were able to train a overparameterized neural network with 35 samples of data in 60 epochs. We attribute this to the effect of having so-called `physical` gradients from training through a PDE solver \cite{holl_physical_2021}. 

\subsection{Physical Interpretation of the Learning Process}
Figure \ref{fig:loss} implies that the training process did indeed minimize eq. \ref{eq:loss}. To understand how this quantity was minimized, it is important to diagnose the simulations that minimize this quantity. To better understand the simulation results, we look at the electric field with respect to time, and find that the wavepacket survives for a much longer duration without damping.

\subsubsection{Electric Field}
\begin{figure}[h]
    \begin{subfigure}{0.5\textwidth}
        \includegraphics[width=\textwidth]{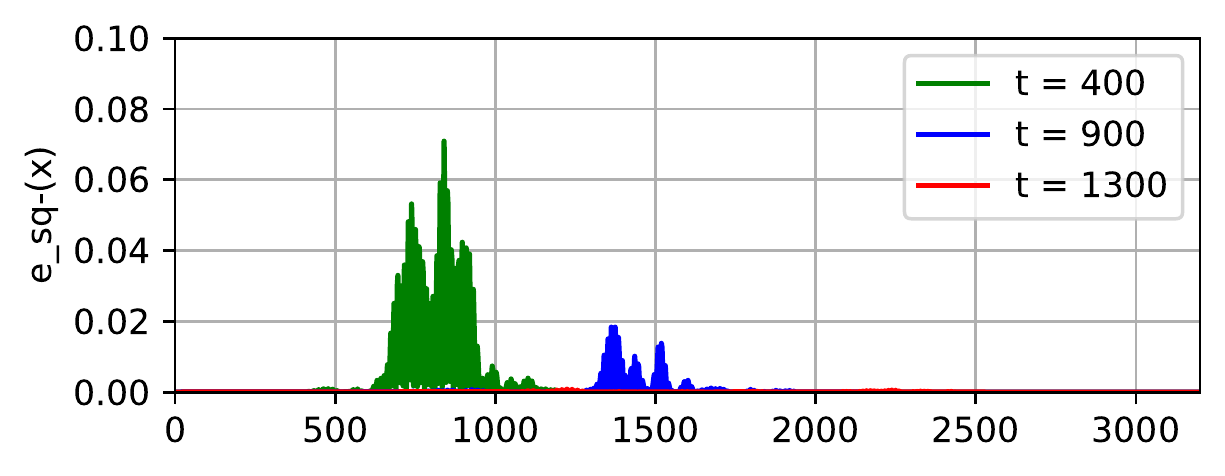}
        \caption{First wavepacket only}
        \label{fig:fld1}
    \end{subfigure}
    \\
    \begin{subfigure}{0.5\textwidth}
        \includegraphics[width=\textwidth]{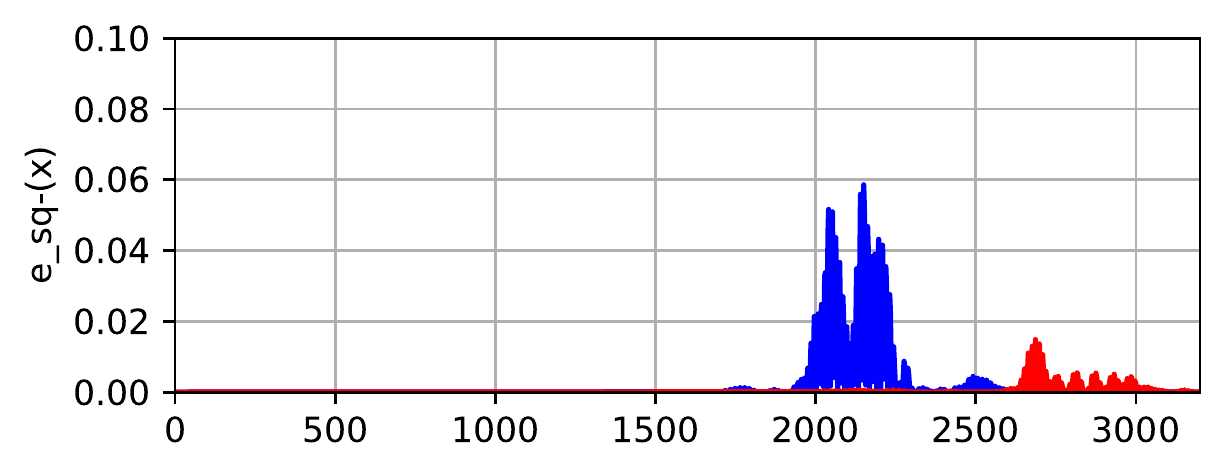}
        \caption{Second wavepacket only}
        \label{fig:fld2}
    \end{subfigure}
    \\
    \begin{subfigure}{0.5\textwidth}
        \includegraphics[width=\textwidth]{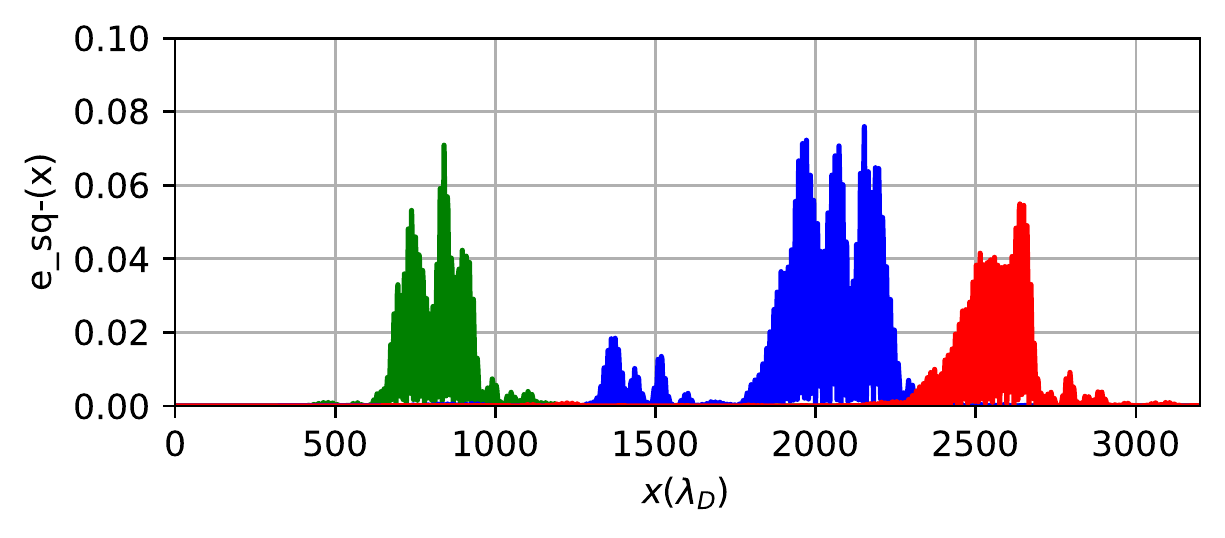}
        \caption{Both wavepackets}
        \label{fig:fld3}
    \end{subfigure}
    \caption{Early in time (green), (c) is the same as (a). Later in time, $t=900 \omega_p^{-1}$ (blue), (a) and (c) show very similar magnitudes for the first wavepacket near $x=1500 \lambda_D$ but the second wavepacket excitation is larger in (c) than (b). At $t=1300 \omega_p^{-1}$ (red), it is clear that (c) is not a superposition of (a) and (b) because (b) has damped away, while (c) retains electrostatic energy suggesting the involvement of a superadditive process.}
    \label{fig:llived}
\end{figure} 

\begin{figure*}[ht]
    \begin{subfigure}{\textwidth}
    \includegraphics[width=\textwidth]{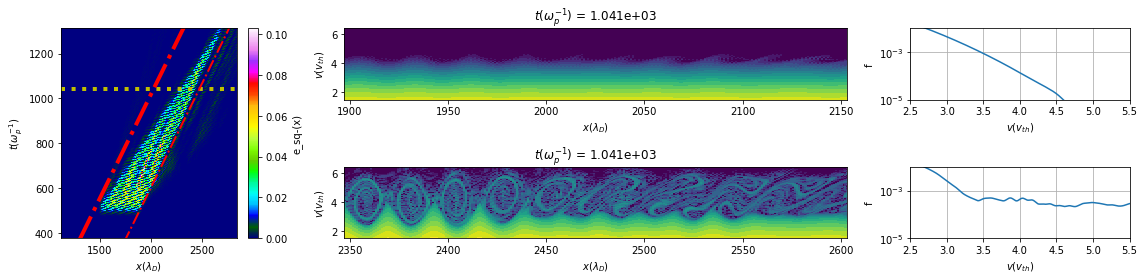}
    \caption{2nd pulse only}
    \label{fig:etch-f}
    \end{subfigure}
    \\
    \begin{subfigure}{\textwidth}
    \includegraphics[width=\textwidth]{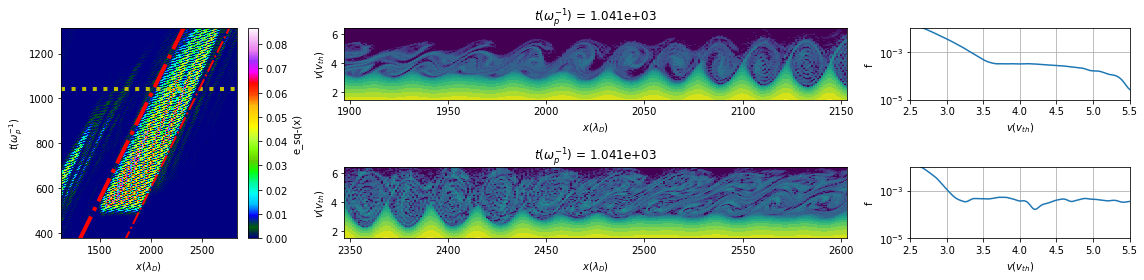}
    \caption{Both pulses}
    \label{fig:llived-f}
    \end{subfigure}
    \caption{\textbf{Left} - Spacetime plot of the electrostatic energy shows the long-lived wavepacket in (b) where the field in (b) survives for a longer duration than in (a). The horizontal line indicates the timestamp of the snapshots in the middle and right. The diagonal dashed-dot lines indicate the spatial location of the snapshots in the middle and right. \\ \textbf{Middle} Phasespace plots at the back (top) and front (bottom) of the wavepacket. In (b), the phase space shows significant activity at the back of the wavepacket while in (a), the distribution function is nearly undisturbed. \\ 
    \textbf{Right} The spatially averaged distribution function. This confirms the fact that the distribution function has returned to a Maxwell-Boltzmann at the back of the wavepacket in (a), while in (b), the distribution function remains flat at the phase velocity of the wave. This is the reason behind the loss of damping.}
    \label{fig:f}
\end{figure*} 

In this section, we analyze the observed behavior of the electric field in the simulations that minimize the metric function. 

Figure \ref{fig:llived} shows the electric field profile for three different simulations. In fig. \ref{fig:fld1}, only the first wavepacket is excited, and in fig. \ref{fig:fld2}, only the second wavepacket is excited. In fig. \ref{fig:fld3}, both wavepackets are excited. 

Early in time, $t=400 \omega_p^{-1}$ (green), when only the first wavepacket has been excited figs. \ref{fig:fld1} and \ref{fig:fld3} agree perfectly. The second wavepacket is excited at $t=500 \omega_p^{-1}$. 

At $t=900 \omega_p^{-1}$, once some time has passed after the excitation of the second wavepacket, the first wavepacket has not fully damped away. It is visible as small bumps in figs. \ref{fig:fld1} and \ref{fig:fld3}. The second wavepacket is also present at this time and easily seen in fig. \ref{fig:fld2}. A larger amplitude wavepacket is seen in fig. \ref{fig:fld3}.

Late in time, the difference in amplitude between the second wavepacket in figs. \ref{fig:fld2} and \ref{fig:fld3} is obvious. The second wavepacket has nearly damped away in fig. \ref{fig:fld2}. In fig. \ref{fig:fld3}, the second wavepacket continues to persist, at nearly the same energy as it was at $t=900 \omega_p^{-1}$.

This superadditive behavior, where $f(x) + f(y) \leq f(x+y)$, suggests that there is a constructive interference. We observe this effect for all wavenumbers we model.

\subsubsection{Phase Space}
To determine the mechanism behind this phenomenon, we turn to the phase-space dynamics. In fig. \ref{fig:f}, (i) is a space-time plot of the electric field. The two dashed-dot red lines at the front and back of the wavepacket are parallel and indicate the velocity of the wavefront. In fig. \ref{fig:etch-f}(i), the front of the wavepacket propagates at a seemingly faster rate than the rear. This is due to the etching effect \cite{fahlen_propagation_2009}. In fig. \ref{fig:llived-f}, the wave survives for a much longer time, as was also illustrated in fig. \ref{fig:llived}. 

Figure \ref{fig:f} (ii) and (iii) are phase-space plots with their center indicated by the intersection of the horizontal timestamp line, and the dashed-dot line at the rear wavepacket red line at the back of the wavepacket. (iv) and (v) correspond to the intersection with the dashed-dot line at the front. (ii) and (iv) show the phase space within a window in x, while (iii) and (v) are the spatially averaged distribution function. (iii) and (v) serve as a proxy for approximating the propensity of Landau damping in that region.

In fig. \ref{fig:etch-f}(ii) and (iii), we see that the rear of the wavepacket is Maxwellian. As previously shown, this is why the rear of the wavepacket damps faster than the front as in fig. \ref{fig:etch-f}(i)  \cite{fahlen_propagation_2009}. 

In the simulations described here, fig. \ref{fig:llived-f} show that the distribution function at the back of the wavepacket has trapped particle activity (fig. \ref{fig:llived-f}(ii)) and near zero slope at the phase velocity of the wave (fig. \ref{fig:llived-f}(iii)). Both plots show that the slope is negligible because of the arrival of streaming detrapped particles from the first wavepacket. Due to this effect, the remergence of Landau damping that occurs due to the loss of trapped particles in isolated wavepackets no longer occurs here. This results in a reduction of the etching and the wavepacket propagates freely for some time while the particles from the first wavepacket propagate and arrive at the rear of the second wavepacket.

\section{Conclusion}
In this preprint, we show how one may be able to discover novel physics using differentiable simulations by posing a physical question as an optimization problem. This removes the physicist from the loop and allows the simulation process to use physical gradients in order to arrive to the solution. The job of the physicist is abstracted one level up and now consists of providing their domain expertise in 1/ determining which parametric or functional dependencies to learn (and whether they should be learned using neural networks or other parameterized functions) and 2/ determining the nature of the objective function that best fulfills the desired criteria.

In fig. \ref{fig:workflows}, we show how one may adapt an existing computational science workflow to the autodidactic process described here. In the work performed here, this process enabled the discovery of parameters in a 4D search space with known bounds but an unknown resolution. We trained the model over a coarse grid in $k_0$ and $t_1$, and learned functions for $x_1$ and $\omega_1$. Using gradient descent here allows an escape from the curse of dimensionality and reduces the problem from a 4D search to a 2D search + 2D gradient descent.

This discovery process is not limited to differentiable simulations. While in fig. \ref{fig:workflows}, $\mathcal{V}$ represents a PDE solve, it only needs to be a AD-enabled function that is a model for a physical system. For example, rather than a PDE solve, $\mathcal{V}$ could represent a pre-trained neural-network-based emulator for experimental data. In such a scenario, one may be able to learn forcing function parameters for an experiment using the proposed workflow.

Finally, in neural network literature, the gradient required for the update is $\partial \mathcal{S} / \partial \theta = \partial \mathcal{S} / \partial \mathcal{G} \times \partial \mathcal{G} / \partial \theta$. We see that this is the same as eq. \ref{eq:nnloss} after the addition of one more node in the computational graph for $\mathcal{V}$, the function that models the physical system. This node, here the PDE calculation, allows the neural network training process to become unsupervised and data-efficient. 

\section*{Acknowledgements}
A. J. wishes to acknowledge D. Strozzi for early and continued encouragement, S. Hoyer for introducing him to JAX, M. Poli for the perspective of a ML researcher, and B. B. Afeyan, W. B. Mori, and B. J. Winjum for discussions on non-linear electron plasma waves.  

A. J. also gratefully acknowledges travel support from Syntensor Inc. 

The authors thank the anonymous reviewers for valuable feedback towards generalizing the content for a non-plasma-physics audience. 




\bibliography{bib}
\bibliographystyle{icml2022}

\newpage
\appendix
\onecolumn

\section{Simulation and Modeling details}

\subsection{Training and Simulation Details}
We vary the independent variables such that 
\begin{align}
    k_0 &\in [0.26, 0.27, ..., 0.32], \\
    t_1 &\in [400, 500, ..., 800],
\end{align}
giving an input space of 35 samples from which we seek to learn these functions. 

\subsubsection{Architecture Details}
We use a neural network with 2 hidden linear layers with 8 nodes activated with a ReLU function. The final layer is activated with a $\tanh$ function. The output is normalized such that $p_1 = p_{norm} \times p_{out} + p_{shift}$ where $p_{out}$ is the output from the $\tanh$ function. We normalize the inputs between 0 and 1, and outputs with reasonable windows. For $x_1$, we allow the entire domain, and $\Delta \omega_1 \in [-0.06, 0.06]$. 

We use the ADAM optimizer with a learning rate of 0.05. 

\subsubsection{Simulation Details}
The training simulations are performed with $N_x = 6656, N_v = 160, N_t = 1200, t_{max} = 1100 \omega_p^{-1}, x_{max} = 6600 \lambda_D$. The amplitude of both drivers is $0.05$. The spatial width is $400 \lambda_D$ and temporal width is $50 \omega_p^{-1}$. 

We solve the Vlasov equation using a sixth order integrator in time \cite{Casas2017} with operator splitting. Both operators are computed with exponential integrators with spectral discretizations in phase space. We use a spectral solver for Gauss's Law.  

In order to more closely match realistic plasma conditions, we also implement a Fokker-Planck collision operator and use a non-zero collision frequency $\nu_{ee} = 10^{-4}$ for these simulations.

We implement absorbing boundaries by increasing the collision frequency of the Krook operator in a small localized region at the boundary \cite{Strozzi2007}.

We reproduce the solvers and tests implemented in \cite{joglekar_vlapy_2020} using JAX \cite{bradbury_jax_2018} and Haiku \cite{hennigan_haiku_2020} to allow the usage of automatic differentiation (AD). The validation tests cover Gauss's Law, Landau damping, and the energy and momentum conservation of the Fokker-Planck implementations when applicable depending on the operator.

\subsection{Validating the Differentiable Simulator by Recovering Known Physics}
We describe an additional test here which involves recovering the linear, small-amplitude electrostatic resonance using the gradient-based implementation enabled by this AD-capable implementation. This ensures that the gradients given by the AD system are physically reasonable and are representative of physical phenomenon.

\subsubsection{Electrostatic Waves}
A fundamental wave in plasma physics is the electrostatic wave in unmagnetized plasmas. The dispersion relation is given in numerous textbooks e.g. \cite{Bellan}, and reproduced here as 
\begin{align}
    1+\frac{\omega_e^2}{k^2} \int dv \frac{dg(v) / dv}{\omega - k v} = 0, \label{eq:edisp}
\end{align}
where $\omega_e$ is the plasma frequency, $k$ is the wavenumber, $\omega$ is the resonant frequency, and $v$ is the independent variable representing the velocity in the integral. $g(v)$ is the distribution function of the plasma particles. This equation has been solved numerically and a lookup table for $\omega$ as a function of $k$ is provided in \cite{Canosa1973}. 

We test the capability of our differentiable simulator framework by reproducing those calculations. To do so, we implement the functionality in fig. \ref{fig:param}. We choose a loss function that minimizes free energy and maximizes the entropy given by
\begin{align}
    \mathcal{C}(\mathbf{x_f}) = -\sum_x (E_x^2 + \sum_v (f~ \log(f) - f_{MX} ~\log(f_{MX})) ~ dv)  . \label{eq:loss}
\end{align}
Because collisions return the distribution to a Maxwell-Boltzmann distribution, it is not enough to simply maximize $f\log(f)$ if one is interested in retaining non-linear effects. Because of this, the entropy in term quantifies the deviation of the distribution function from a Maxwell-Boltzmann distribution. 

In this test, we launch plasma waves using a ponderomotive driver in a box with periodic boundary conditions. For each optimization, the wavenumber $k$ and the box size $x_{max}$ change. We provide a lower bound of $0.5$ and an upper bound of $1.5$ to the optimization algorithm. We use the L-BFGS algorithm implemented in SciPy \cite{scipy_10_contributors_scipy_2020}. We set $f_\text{tol}=r_\text{tol}=10^{-12}$.  We learn the $\omega$ that will minimize the value given by eq. \ref{eq:loss} and ensure that it is within 2 decimal places of the previously computed numerical solution of eq. \ref{eq:edisp} \cite{Canosa1973}.




\end{document}